\documentclass[prb,aps,preprint,showkeys]{revtex4}
\usepackage{amsmath}
\usepackage{times}
\usepackage[dvips]{epsfig}
\newcommand{\Kaption}[2]{\noindent FIG.~{#1}:~{#2}\\$\mbox{}$\\}
\begin{document}
\title{Charge localization in multiply-charged clusters and their electrical properties: Some insights into electrospray droplets}
\author{David A. Bonhommeau\footnote{Corresponding author: david.bonhommeau@univ-reims.fr}}
\affiliation{GSMA CNRS UMR7331, Universit{\'e} de Reims Champagne-Ardenne, 
UFR~Sciences Exactes et Naturelles, Moulin de la Housse BP~1039, 51687 Reims Cedex 2, France}
\author{Riccardo Spezia, Marie-Pierre Gaigeot\footnote{Corresponding author: mgaigeot@univ-evry.fr}\footnote{Institut Universitaire de France, 103 Blvd St Michel, 75005 Paris, France}}
\affiliation{LAMBE CNRS UMR8587, Universit{\'e} d'Evry val d'Essonne, Blvd F.~Mitterrand, B{\^a}t Maupertuis, 91025 Evry, France}

\date{\today}

\begin{abstract}
The surface composition of charged Lennard-Jones clusters A$_N^{n+}$, composed of $N$ particles ($55\le N\le 1169$) among which $n$ are positively charged 
with charge q,  thus having a net total charge $Q=nq$,  
is investigated by Monte Carlo with Parallel Tempering (MCPT) simulations.
At finite temperature, the surface sites of these charged clusters are found to be preferentially occupied by charged particles carrying large charges, due to Coulombic repulsions, but the full occupancy of surface sites is rarely achieved for clusters below the stability limit defined in this work. 
Large clusters ($N=1169$) follow the same trends, with a smaller propensity for positive particles to occupy  the cluster surface at non-zero temperature.
We show that these charged clusters rather behave as electrical spherical conductors for the smaller sizes ($N\le 147$) but as spheres uniformly charged in their volume for the larger sizes ($N = 1169$). 
\end{abstract}


\maketitle
\section{Introduction}

The question of ion localization at interfaces has aroused considerable interest in the past decade, from the study of salts, acids, and bases at liquid interfaces\cite{SGopalakrishnan2006}, to nanodroplets\cite{SGopalakrishnan2006}, atmospheric aerosols\cite{BJFinlayson2009}, and hydrophobic droplets\cite{RVacha2011}.
The affinity of ions for the very surface of bulk water is a subtle topic, where the specific properties of ions and their effect on the structure of the solvent come into play\cite{BCGarrett2004,TMChang2006,RVacha2011,CCaleman2011}.  
However, in finite-sized charged clusters or droplets, the distribution of charges is further affected by the fact that they are confined to a roughly spherical volume.
In such systems, one issue is the competition between charge localization inside or at the surface. 
It is this latter effect that we study in the present work, employing a model whose liquid structure is far simpler than that of water.
One of the most famous theoretical problem related to this question is the determination of the configuration that minimizes the Coulombic energy of $N$ point charges located on a sphere, 
as initially proposed by J. J. Thomson. 
Such a fundamental problem finds applicability in modern investigations such as the formation pattern of virus capsids\cite{CJMarzec1993},  fullerene stability\cite{HWKroto1985}, or  multielectron bubbles at helium interfaces\cite{PLederer1995}, and has been treated by a number of theoretical approaches among which simulated annealing\cite{LTWille1986}, genetic algorithms\cite{JRMorris1995}, basin-hopping\cite{DJWales2006,DJWales2009}, and constrained global optimization\cite{ELAltschuler1994}. 

Electrospray Ionisation (ESI) in Mass Spectrometry\cite{JBFenn2003} is a domain where charged droplets are formed and some interrogations on the charges localization still persist.
In these experiments, charged droplets composed of solvent and analyte molecules (typically biomolecules) are produced at finite temperature.  
After several cycles of evaporation, the analyte carrying part of the total droplet charge may eventually be ejected from the droplet and be analysed by mass spectrometry techniques in the gas phase.
Beside the open question on the fragmentation mechanisms of these charged droplets\cite{MDole1968,JVIribarne1976,PKebarle2009b}, one important issue is the localization of the embedded charges, {\it i.e.} within or at the surface of the droplets, and the influence of this charge localization on the droplet stability. 
Recent atomistic simulations have shown that the common belief that positive charges are located at the droplet surface should be taken with caution\cite{EAhadi2010,IMarginean2006}.
Localization of the embedded charges within charged droplets and the subtle effects on charges delocalization are the subjects tackled in the present investigation. 

Inspired by the work of Jortner and co-workers, we have modeled  ESI droplets as charged clusters\cite{ILast2005} with a simple and generic representation of the interactions between individual particles. 
In the rest of this  work, 
droplets containing a positive net total charge will be considered,  
thus corresponding to positive-ESI in experiments. 
Conclusions will be transferable to negative-ESI mode. 
In the positive-ESI mode, droplets with a net total positive charge are mainly produced 
because of the voltage applied in the capillary interface that drives positively 
charged entities towards the vacuum chamber. 
This does not  preclude that negative charged particles are present 
within the droplet, up to the point that pure water droplets electrosprayed 
in positive-ESI mode can form negatively charged droplets, 
and the reverse, i.e. pure water droplets electrosprayed 
in negative-ESI mode provide a majority ($>$70\%) of positively charged droplets,
as shown by Jarrold and coworkers~\cite{JTMaze2006,LWZilch2008,LWZilch2009}. 
Note also that the droplets are electrosprayed from a mixture of  solvents (water, methanol and their mixture are typically  employed in ESI of biomolecules), biomolecules and counterions (salts and buffers from the initial solution), for typical applications in proteomics and genomics. 
Residual  counterions are thus  present in the final positively charged droplets formed. 
Other sources for the presence of negative charges within a positively charged 
electrosprayed droplet are typically collisions with metal surfaces 
(from the walls of 
the capillary interface and skimmers for instance) or  
electric fields above the Taylor limit~\cite{RLGrimm2005}. 
These remarks convey the idea that positively charged droplets of interest 
to our present modeling will be composed of charged particles carrying a 
positive charge $q>0$, but it is not unreasonable to model charged droplets 
with a total net positive charge carrying both positive $q>0$ and negative $q<0$ 
individual charges. 

In the present investigation, we model charged droplets as A$_N^{n+}$ clusters 
composed of $N$ individual particles among which $n$ are charged particles, each one  
carrying the charge $q$ ($q>0$ in most calculations, but we will also investigate 
charged clusters composed of both $q>0$ and $q<0$ charges). 
The cluster has a total net positive charge $Q=nq$. 
We investigate the influence of structural changes (ie, cluster size $N$, number of embedded charged particles $n$) as well as electrostatic  (ie, charge sign and value, polarisability) and thermodynamic (ie, temperature) changes on the propensity for the  charged particles to occupy the surface of the charged A$_N^{n+}$ clusters. 
We are interested in A$_N^{n+}$ clusters composed of $55\le N\le 1169$ particles. 

The present investigation is based on Monte Carlo with Parallel Tempering (MCPT) simulations.
We use a simplified dimensionless form of the interaction potential between the particles, that could be qualified in modern language as a mesoscopic highly-coarse-grained representation. 
It is based on  Lennard-Jones and Coulomb pairwise interactions between individual spherical particles, completed by the addition of $N$-body polarisation. 
The interactions are written in dimensionless units. 
Each individual coarse-grained particle is either neutral ($q=0$) or charged ($q\ne 0$),  
and may carry an induced dipole moment if polarisation effects are taken into account in the calculation.
Within this model, one coarse-grained particle can be seen as either one biomolecule (for instance one amino acid, one peptide, or even one protein, for the proteic family), or as one coarse-grained solvent particle mimicking a few  microscopic solvent molecules. 
In this latter case, our representation follows the coarse-grained models of water 
developed in the recent literature, where one chargeless coarse grain can be mapped to up to  four individual microscopic water molecules. 
See for instance the review by Klein {\it and coll.}~\cite{XHe2010} and 
the popular Martini coarse-grained model~\cite{LMonticelli2008}. 
The neutral spherical particles in our model can thus be seen as 
coarse grains for the solvent, i.e.  water or methanol (and mixtures of) as typically used in 
ESI of proteins and non-covalent complexes of proteins~\cite{ATIavarone2000,MZhou2010}.  
The charged coarse grains mimic biomolecules, 
and there is no explicit charge transfers between solvent grains and biomolecules grains in the model 
(i.e. the ESI charging process is not directly modeled). 
It is easy to think of such spherical coarse grains in terms of charged 
amino acids (either for the native charged residues or 
after protonation of neutral residues in positive-ESI) or in terms of charged 
folded peptides and globular proteins. 
Similar spherical coarse grains are also used in the literature in order to simulate highly coarse grained non-globular proteins, i.e. patchy proteins~\cite{NKern2003,EGNoya2007,SWhitelam2009}   
where the patches account for the anisotropy of interactions (and thus non spherical shapes). 
Here we consider a simple isotropic coarse-grained model of proteins, but one 
improvement will certainly be to introduce such patches.  
Obviously, the charged grains can also represent ionized water (a grain of water 
including one H$_3$O$^+$ molecule in positive-ESI), or a residual counterion 
(a grain would be seen as the counterion and its immediate solvation sphere). 
Our purpose is not to develop coarse-grained models dedicated to a specific 
solvent or to specific biomolecules. 
Rather we want to use a generic  representation of the interactions between individual grains, and to that end dimensionless units are used. 
This is an advantageous representation as one single simulation can be used for describing several different cases, i.e. encompassing several solvent and analyte compositions, provided an inverse mapping between dimensionless units and real units is eventually done. 

Section~\ref{sec-method} of this paper describes the dimensionless potential energy surface used in the present work and  computational details about the Monte Carlo with Parallel Tempering method.
Results on charge localization and its relation to the excess charge delocalization are reported in Section~\ref{sec-results}.
Concluding remarks are  collected in Section~\ref{sec-conclusion}.   

\section{Method}\label{sec-method}

\subsection{Interaction model}

Our A$_N^{n+}$ clusters (i.e. composed of $N$ spherical particles, $n$ are 
positively charged with the individual charge $q$, total net charge of the cluster $Q=n q$) 
are bound by Lennard-Jones and electrostatic interactions (charge-charge, charge-induced dipole, induced dipole-induced dipole) between individual spherical particles. Polarisation is expressed according to the point dipole model of Thole\cite{BTThole1981,MSouaille2009}.
Defining the usual Lennard-Jones reduced energy $V^*=V/u$ and length $r^*=r/\sigma$, 
where $u$ and $\sigma$ are the Lennard-Jones pair well-depth and diameter parameters, 
the reduced charge and polarizability are respectively $q^*=q/(4\pi\epsilon_0\sigma\,u)^{1/2}$ and $\alpha^* = \alpha/(4\pi\epsilon_0\sigma^3)$, and the reduced dipole moment is $\vec{p}^{\,*}=\vec{p}/(4\pi\epsilon_0\sigma^3\,u)^{1/2}$,  where $\epsilon_0$ is the vacuum permittivity. 
The dimensionless potential used in the present simulations is: 
\begin{equation}
V^*(\mathbf{r}) =\sum_{i,j \atop j>i} 4\left(\frac{1}{r_{ij}^{*12}} - \frac{1}{r_{ij}^{*6}}\right) + \frac{1}{2} \sum_i \vec{p}^{\,*}_i [\alpha^*_i]^{-1} \vec{p}^{\,*}_i + \frac{1}{2}\sum_{i,j \atop i\ne j} (q^*_i +\vec{p}^{\,*}_i\mathbf{\nabla}_i)(q^*_j - \vec{p}^{\,*}_j\mathbf{\nabla}_i)\phi^s(r^*_{ij})
\end{equation}
\noindent 
considering that all spherical particles (or grains) have the same size. 
In this equation, $r^*_{ij}$ is the dimensionless distance between particles $i$ and $j$, $q^*_i$, $\alpha^*_i$ and $\vec{p}^{\,*}_i$ are respectively the dimensionless charge, polarisability and induced dipole of particle $i$. 
$\phi^s(r^*_{ij})$ is the Thole damping function
\begin{equation}
\phi^s(r^*_{ij}) = \frac{1}{r^*_{ij}}\left[1-\left(1+\frac{a\,\gamma_{ij}}{2}\right)e^{-a\,\gamma_{ij}} \right]
\end{equation}
with $a = 2.1304$ and $\gamma_{ij} = r^*_{ij}/(\alpha^*_i\alpha^*_j)^{1/6}$.
The dimensionless polarisability is chosen equal to $\alpha^*_i=0.1$ for all the polarisable particles and the charge  $q^*_i = q^*$ is the same for all the charged particles, unless specified otherwise (ie, when we consider droplets with positive and negative charged particles).

\subsection{Monte Carlo with Parallel Tempering (MCPT) simulations}

The MCPT simulations were performed in spherical containers of radii $R = 4.5\sigma$, $5.0\sigma$, $5.5\sigma$, and $10\sigma$ respectively for the cluster sizes $N=55, 100, 147$, and $1169$  (note that these values exceed the typical radii used for modelling the corresponding neutral clusters so as to provide more flexibility for geometrical distortion of charged clusters). 
The centre of mass of the cluster and the container centre are constrained to coincide.

Each parallel MCPT simulation is a set of $N_{rep}$  Monte Carlo simulations performed at different dimensionless temperatures $T^*_i$ with different initial random seeds $\xi_i$.
The temperatures $T^*_i$ encompass the temperature range from the solid-state (low-temperature asymptote of the heat capacity curves) to liquid-state temperatures (high-temperature asymptote of the heat capacity curves). 
We adjusted the temperature scale for each investigated cluster in order to get the better compromise between resolution of the melting peak and convergence of the simulation,  maintaining the required overlap between adjacent energy distributions between 40\% and 90\%.\cite{NoteMCPT}

Three MC moves have been adopted in the MCPT simulations: (i) displacement of particles, (ii) swap between a neutral and a charged particle, (iii) swap between configurations from adjacent replicas at temperatures  $T^*_i$ and $T^*_{i\pm 1}$.
Simulations were performed for $10^6$ to $10^8$ sweeps (one sweep = $N$ MC steps) depending on the cluster size and whether polarisation is taken into account (more time consuming calculations). 
This number of MC steps and thus the convergence of the MCPT calculations have been checked on the heat capacity curves of neutral clusters, comparing our results to available data from the literature.\cite{JPKDoye1998,EGNoya2006}
Global geometry optimisations have also been performed with the GMIN code\cite{DJWales1997}.

\subsection{Stability criterion}
Despite the cohesion due to Lennard-Jones interactions and polarisation, charged clusters should become rapidly unstable as the total reduced charge $Q^*=nq^*$ increases. 
We therefore defined a stability limit below which clusters can be regarded as stable and are included and considered in our analyses. 
The stability criterion is established at the melting temperature, and our 
definition  is based on the fraction of configurations in which evaporation is detected. An individual configuration is considered to undergo evaporation if at least one particle reaches the container edge within a small distance of $1/2$ in reduced units (or $\sigma/2$ without reduced units). 
For $N\le 147$, the neutral clusters are not prone to evaporation at the melting temperature and a charged cluster ($T^*_{melt}$ roughly in the range 0.23-0.36 for the charged clusters considered here) was considered stable if no evaporation occurred at this temperature among at least 90\% of the configurations explored by the MCPT simulation. 
For $N=1169$, the neutral cluster is already prone to some evaporation (in $\sim 20-25\%$ of the configurations) at the melting temperature and a charged cluster ($T^*_{melt}$ roughly in the range 0.52-0.55 for the charged clusters considered here) was considered stable if no evaporation occurred at this temperature among at least 60\% of the configurations explored by the MCPT simulation.

\section{Results}\label{sec-results}

\subsection{Localization of charged particles and temperature effects} \label{localization}

\begin{figure}[t]
\epsfig{file=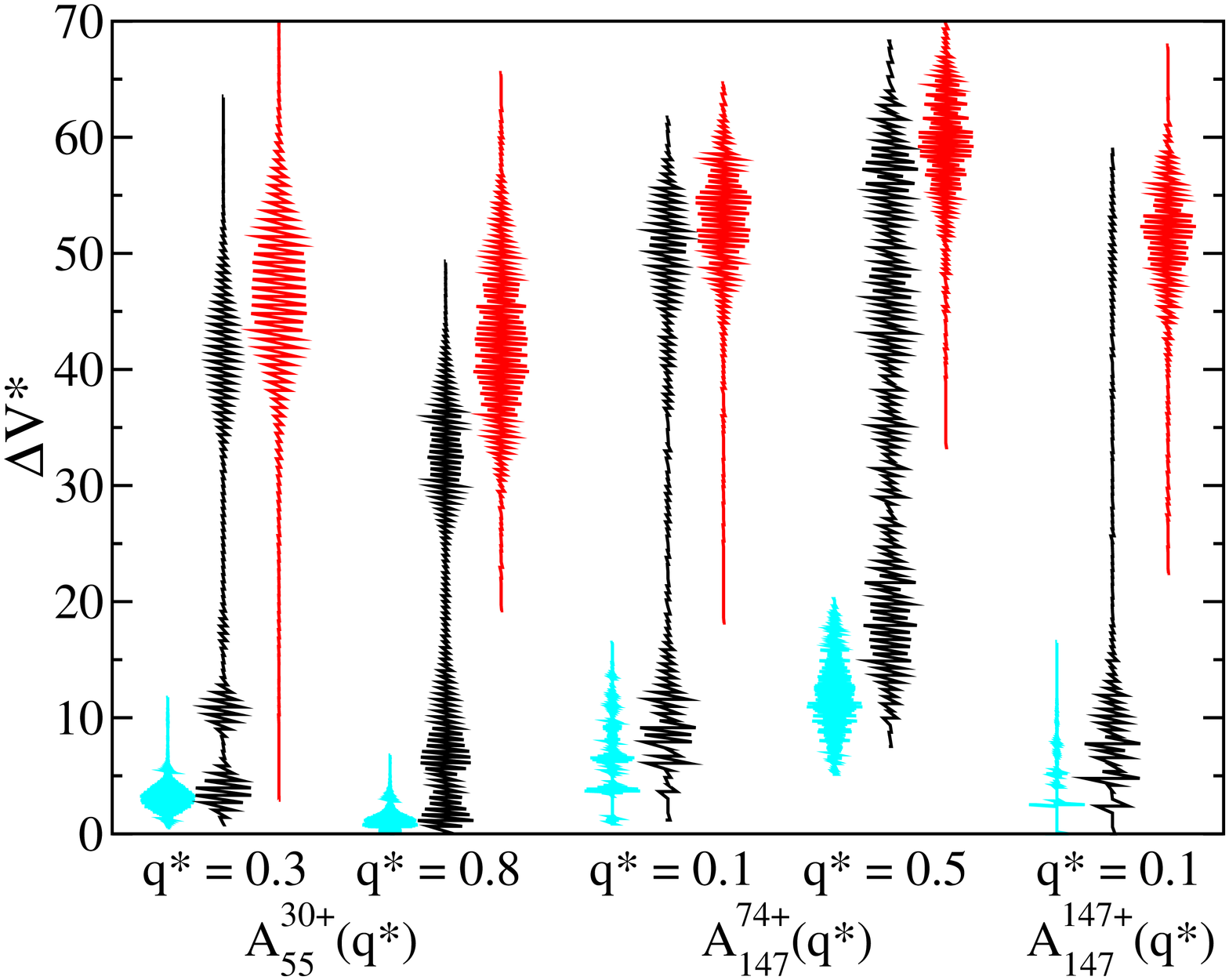,width=0.35\textwidth,angle=270}
\caption{}
\label{fig-LMIN}
\end{figure}

At $T^*=0$, the charged particles of A$_N^{n+}$ clusters are located at the cluster surface.
For instance, the global minimum configurations of icosahedral $A_{55}^{12+}$ and $A_{147}^{12+}$ clusters have their twelve charged particles located on the twelve icosahedron vertices whatever the value of the charge $q^*$.
However, when the temperature increases, the global minimum configuration is (not unexpectedly) not systematically entropically favoured.
This can be seen in Figure~\ref{fig-LMIN} where the energy difference $\Delta V^*$ between the local minima belonging to the basins explored during the MCPT simulations and the global minimum is plotted for selected small clusters at solid-state (cyan), melting (black), and liquid-state (red) temperatures, respectively.
Note that the melting temperature is defined as the temperature at the heat capacity peak, the solid-state and liquid-state temperatures are the minimum and maximum temperatures of our temperature ranges respectively, and each distribution was obtained by performing local minimisations with a conjugate gradient method over at least $10^3$ configurations collected along the MCPT simulations.

At low temperature, the energy distributions of the smaller clusters such as A$_{55}^{30+}$ are located in the immediate surrounding of the global minimum  ($\Delta V^* = 0$, Fig.\ref{fig-LMIN}), all the more when $q^*$ is large.  
Conformations structurally close to the global minimum are explored and the charged particles are thus preferentially located at the cluster's surface, though possibly occupying sites on the faces or edges rather than the icosahedron vertices.
For $N=147$ with $n=74$ embedded charges, the surroundings of the global minimum can only be explored for small $q^*$ values. 
Going from $q^*=0.1$ to $q^*=0.5$, $\Delta V^*$ already amounts to 5--20, and the global minimum energy basin is no longer significantly explored during the MCPT simulation for $q^*=0.5$. 
Fully charged clusters such as $A_{147}^{147+}$ are peculiar cases with much simpler energy landscapes, preserving the  icosahedron structure of the neutral cluster, and therefore exploring very few minima other than the global one. 
At liquid-state temperatures (red plots in Fig.\ref{fig-LMIN}), the global minimum surroundings are no longer explored, 
and much higher energy conformations are found for all the systems presented in Fig.~\ref{fig-LMIN}. 
At these temperatures, the clusters adopt a variety of distorted geometries in order to maintain their ability to accomodate, to some extent, charged particles at their surface as discussed below.

\begin{figure}
\epsfig{file=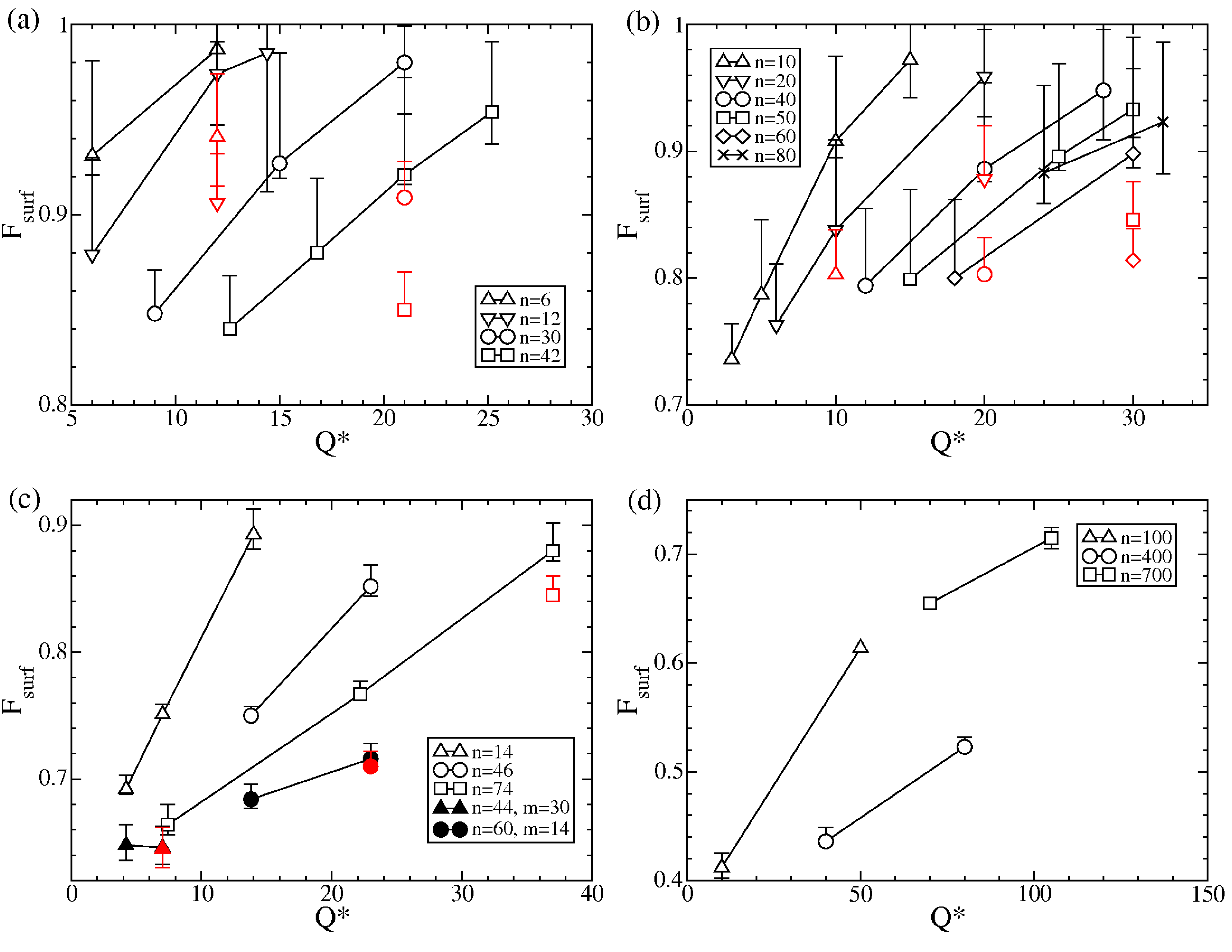,width=1.0\textwidth,angle=0}
\caption{}
\label{fig-Fsurf}
\end{figure}

We have plotted in Figures~\ref{fig-Fsurf}a-d the average fraction $F_{surf}$ of charged particles located at the cluster surface as a function of the total cluster charge $Q^*$ ($Q^* = nq^*$ for A$_N^{n+}$,  $Q^*=(n-m)q^*$ for A$_N^{n+m-}$), at the melting temperature of the charged clusters. 
The fraction $F_{\rm surf}$ is defined by the ratio of the number of charged particles located at the surface over the maximum number of charged surface particles (the method for surface particle definition is described in Ref.\cite{MAMiller2012}). 
A value of $F_{\rm surf} = 1$ would correspond to charged particles only located at the cluster surface. 
Note that the number of surface sites for neutral A$_{55}$, A$_{100}$, A$_{147}$, A$_{1169}$ clusters are respectively 42, 69, 92, and 449 at $T^* = 0$.
These values are subject to a 5-10\% increase at the highest temperatures of the MCPT simulations.

We first discuss  the results obtained without polarisation for the smaller clusters, ie $N \leq 147$ (black open symbols in Fig.~\ref{fig-Fsurf}a-c). 
For these systems, $F_{\rm surf}$ varies between 0.66 (see Fig.~\ref{fig-Fsurf}c) and 0.99 (see Fig.~\ref{fig-Fsurf}a), at  the melting temperatures. 
The propensity of the $n$ positive particles to occupy the surface of A$_N^{n+}$ clusters steadily increases as $Q^*$ increases. 
This is due to the increase in the repulsive Coulomb interactions that repel the charges towards the cluster surface 
as $q$ increases. 
$F_{\rm surf}$ sensitively depends on the repartition $(n, q^*)$ for a given cluster size $N$.  
Two clusters with equal $Q^*$ such as A$_{55}^{30+}(q^*=0.7)$ and A$_{55}^{42+}(q^*=0.5)$ [$Q^*=21$], or A$_{100}^{10+}(q^*=1.0)$ and A$_{100}^{20+}(q^*=0.5)$ [$Q^*=10$], or A$_{147}^{14+}(q^*=1.0)$ and A$_{147}^{46+}(q^*=0.3)$ [$Q^*\approx 14$], present different fractions of charged particles at the cluster surface: 
for a given $Q^*$, stable clusters with a larger number of charged particles $n$ have smaller charge values $q^*$ and therefore display smaller fractions of charged particles occupying the cluster surface sites.
The upward error bars in Figs~\ref{fig-Fsurf}a-b correspond to $F_{\rm surf}$ values spanned at the lower temperatures of the MCPT simulations (solid-state). 
It shows that  $F_{\rm surf}$ increases by $5-10\%$ at solid-state temperatures, possibly reaching unity  for certain cases of $N=55$ and $N=100$. 
Temperature is thus a pivotal parameter on which the fraction of charged particles located at the cluster surface strongly depends. 

Figure~\ref{fig-Fsurf} also reports $F_{\rm surf}$ values when including polarisation, showing a slight decrease in the number of charged surface particles.
For instance, $F_{\rm surf}$ decreases from 0.974 to 0.906 for A$_{55}^{12+}(q^*=1)$, or from 0.880 to 0.845 for A$_{147}^{74+}(q^*=0.5)$. 
This loss of charged surface particles is significantly reduced for mixed clusters 
(ie, composed of both positive and negative charged particles).
For instance, $F_{\rm surf}$ only decreases from 0.716 to 0.710 for A$_{147}^{60+14-}(q^*=0.5)$.

For $N=1169$, the general trends observed on small clusters, \textit{i.e.} increase of $F_{\rm surf}$ with $Q^*$ at fixed $n$ and faster increase of $F_{\rm surf}$ for small $n$ at fixed $Q^*$, are preserved.
However, unlike smaller clusters, a smaller fraction of the charged particles are located at the surface of the A$_{1169}^{n+}$ clusters. 
The ratio of surface sites compared to inner sites is lowered in larger clusters, and the full occupancy of surface sites is never achieved although $n$ may greatly exceed the number of accessible surface sites. 
For instance,  the number of surface sites of A$_{1169}^{700+}$ clusters is about $450-490$ over the temperature range spanned in the MCPT simulations but the cluster surface is never fully occupied ($\sim$ 72\% maximum occupancy found for the cases investigated here). 

It is also relevant to explore whether this surface propensity for the 
positively charged particles remains when both positive and negative charged particles 
are embedded within the cluster. 
We have investigated that point in the case of the prototype 
A$_{147}^{44+30-}$ and A$_{147}^{60+14-}$ clusters, 
both composed of a total of $74$ embedded charged particles with fixed $|q^*|$, 
respectively with 44 $q^*>0$ charges and 30 $q^*<0$ charges (A$_{147}^{44+30-}$, $Q^*=+14|q^*|$) 
and 60 $q^*>0$ and 14  $q^*<0$ charges (A$_{147}^{60+14-}$, $Q^*=+46|q^*|$).
These mixed clusters have therefore the same total charge as A$_{147}^{14+}$ and A$_{147}^{46+}$ respectively, and the same number of charged particles as A$_{147}^{74+}$, but their fraction of charged particles on the cluster surface is significantly smaller (see Fig.~\ref{fig-Fsurf}c, filled black symbols). 
The negative particles are more prone to reside inside the cluster than the positive particles.
For instance, only one third of the negative particles present in A$_{147}^{60+14-}(|q^*|=0.5)$ clusters are located on the surface. 
Due to attractive Coulombic interactions, the positive particles tend to organise themselves around the negative particles, thus decreasing their presence at the clusters surfaces. 
There is thus a competition between solvating the negative particles within the cluster and reaching the cluster surface. 
As an example, we found that $F_{\rm surf}\approx 0.81$ for A$_{147}^{60+14-}(|q^*|=0.5)$ which is below the fractions obtained for A$_{147}^{46+}(q^*=0.5)$ and A$_{147}^{74+}(q^*=0.5)$, respectively $0.85$ and $0.88$.
Including polarisation does not change these conclusions. 

\subsection{Are charged droplets spherical conductors or uniformly charged spheres? }
In a recent paper, Ahadi and Konermann\cite{EAhadi2010} very nicely showed that although excess charges (Na$^+$ ions in their case) embedded in a water droplet are not located at the droplet surface the excess charge is confined to a thin layer at the droplet periphery, so that the droplet acts as an ideal spherical conductor. 
Their demonstration is based on the electrostatic potential energy between the charged droplet and an external point charge probe. 
The structures of the droplets are  extracted from 
room temperature microscopic molecular dynamics simulations of 10 Na$^+$ ions embedded in a water droplet of roughly 20~\AA \ ($\sim$1250 water molecules), close to the Rayleigh limit.
They found that the dependence of the electrostatic potential energy with respect to the distance $r$ between the probe and the center of mass of the droplet 
is identical to the one obtained for a spherical conductor, 
ie, constant within the radius of the droplet and varying as $1/r$ outside the droplet. 
The argument used in the paper in order to reconcile the apparent 
contradiction between the charges location (inside the droplet) and the 
excess effective charge confined to a thin layer at the droplet surface 
is based on charge-induced orientation of the water dipole moments: 
the water molecules solvate the Na$^+$ ions within the droplet which induces an 
``orientational polarization'' or ``orientational alignment'' 
of the water molecules. This consequently acts in a way so that 
the excess charge is transfered to the droplet periphery. 
The fact that the water molecules are polar molecules that form intermolecular hydrogen bonds is probably pivotal in the final charge-induced orientation of the water dipole moments. 
One might then logically wonder how much of these arguments are transferable and applicable to droplets not composed of explicit microscopic water molecules, 
but instead composed of solvent molecules having neither a permanent dipole moment neither forming 
hydrogen bonds. 
Our model for coarse-grained charged droplets provides an excellent test, and we apply now the same electrostatic calculations as in Ref.\cite{EAhadi2010} on selected charged clusters investigated in Section~\ref{localization}. 
Note that some calculations presented below will include polarization 
effects in terms of induced dipoles on the particles, 
which will account for  ``charge transfers'' between the particles of the clusters (an effect not taken into account in Ref.\cite{EAhadi2010}). 

\begin{figure}
\epsfig{file=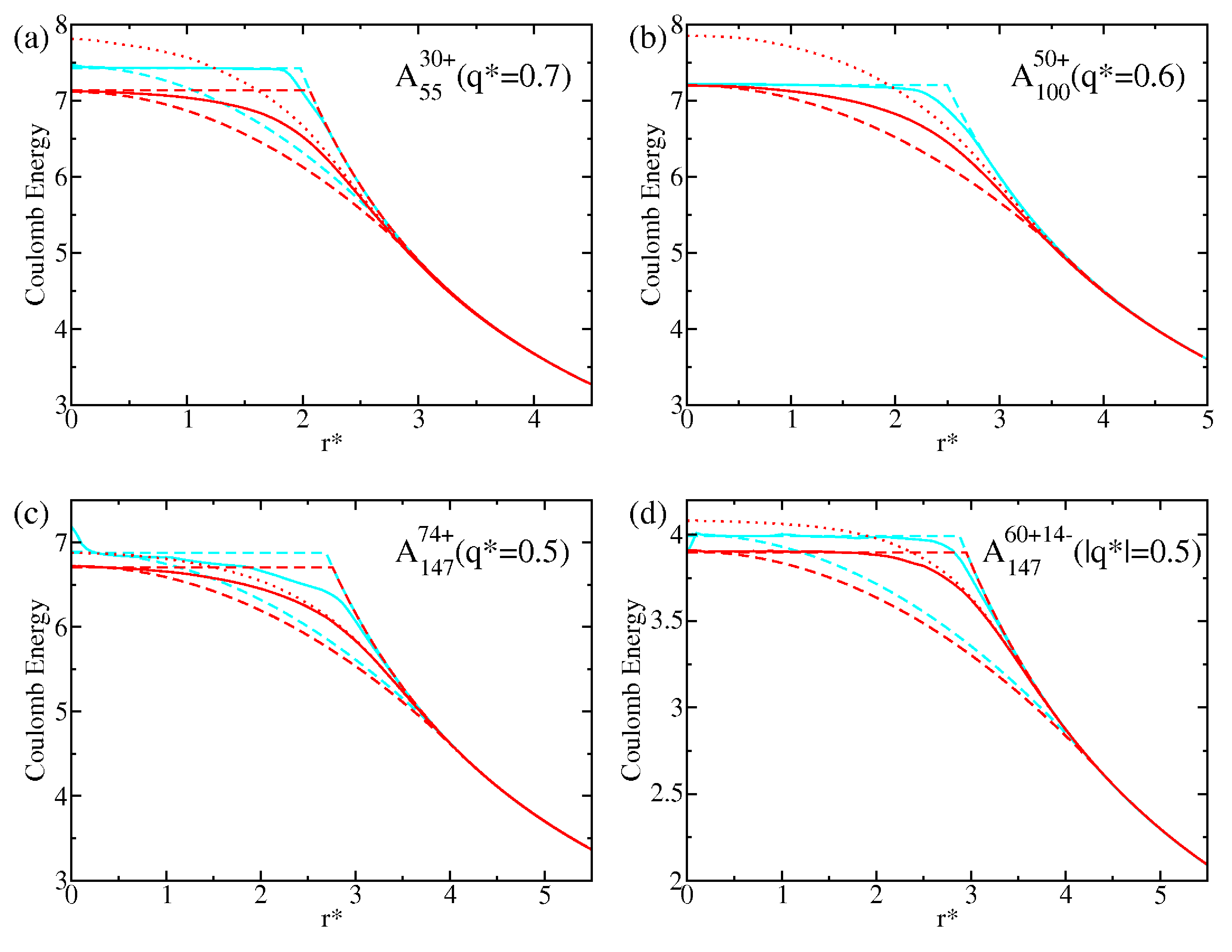,width=1.0\textwidth,angle=0}
\caption{}
\label{fig-Ecoul}
\end{figure}

Following Ahadi and Konermann\cite{EAhadi2010}, we present in Fig.\ref{fig-Ecoul} the Coulomb potential energy $U^*_{MC}$ curves obtained from our MCPT simulations: 
\begin{equation}
U_{MC}^*(r^*) = \sum_{i=1}^N \frac{q^*_{probe}q^*_i}{|\vec{r}_i^{\,*}-\vec{r}^{\,*}|} 
\label{eq:Ucoul}
\end{equation}
between a point charge probe $q^*_{probe} = q^*$ and all $N$ particles $q^*_i = q^*$  of the charged cluster A$_N^{n+}$. 
In this equation, $\vec{r}_i^{\,*}$ and $\vec{r}^{\,*}$ are respectively the positions of particle $i$ and of the probe point charge defined with respect to the cluster center of mass. 
$U^*_{MC}(r^*)$ is averaged over at least $10^3$ Monte Carlo configurations, and 
anticipating that A$_N^{n+}$ clusters are not exactly spherical we have furthermore averaged  over the three directions of space and over three planes ($xy$, $yz$, $zx$). 
In Ref.\cite{EAhadi2010} the droplet charge was close to the Rayleigh limit, we have therefore used some of our results obtained for clusters with a relatively high total charge $Q^*$, although this charge may still be far from the Rayleigh limit.

The electrostatic arguments behind Eq.(\ref{eq:Ucoul}) are now the following. 
If we consider our charged clusters as roughly spherical in shape, 
they might either behave as a spherical conductor charged at the surface 
or  as a sphere uniformly charged in volume. 
Following the Gauss law for the calculation of the radial electrostatic field $\vec{E}^*(r^*)$ and the related electrostatic potential $V^*(r^*)$ created by such objects anywhere in space, i.e. for $r^* < R^*_0$ and  $r^* > R_0^*$ where $R_0^*$ is the radius of the sphere, 
we get the following expressions for the Coulomb potential energy 
(see Appendix for the derivation of the expressions in reduced units): 
\begin{subequations}
\begin{eqnarray}
\label{Eq:idealConductor1}
r^* < R^*_0 & : & U^*_{surf} = 4\pi q_{probe}^*\sigma_Q^*R^*_0\\ 
\label{Eq:idealConductor2}
r^* > R^*_0 & : & U^*_{surf} = \frac{4\pi q_{probe}^*\sigma_Q^* R_0^{*2}}{r^*}.
\end{eqnarray}
\end{subequations}
for the spherical conductor charged at its surface with the surface density of charge $\sigma_Q^*=Q^*/S^*$, 
and 
\begin{subequations}
\begin{eqnarray}
\label{Eq:idealVolume1}
r^* < R_0^* & : & U^*_{vol} = \frac{4\pi q_{probe}^*\rho_Q^*}{2}\left(R_0^{*2}-\frac{r^{*2}}{3}\right)\\ 
\label{Eq:idealVolume2}
r^* > R_0^* & : & U^*_{vol} = \frac{4\pi q_{probe}^*\rho_Q^* R_0^{*3}}{3r^*}.
\end{eqnarray}
\end{subequations}
for the sphere uniformly charged in volume with the volume density of charge $\rho_Q^*=Q^*/V^*$. 
The Coulomb energy between a spherical conductor or a sphere uniformly charged in volume and an external probe particle varies as $1/r^*$ for $r^* > R_0^*$ in both cases.
However, the behaviour largely differs for $r^* < R_0^*$ since this energy is constant (the electrostatic potential is constant) inside the spherical conductor and decreases as $-r^{*2}$ when the sphere is uniformly charged in volume.

%
%
%
%
%
%
%

We have plotted in Fig.~\ref{fig-Ecoul} the Coulomb energies $U^*_{MC}$ extracted from the MCPT simulations (solid curves) at solid-state (cyan curves) and liquid-state (red curves) temperatures. 
We compare these results to the ideal Coulomb energies (dashed curves, same convention of colours as above for solid/liquid states) $U^*_{surf}$ defined by Eqs.~(\ref{Eq:idealConductor1})-(\ref{Eq:idealConductor2}) and $U^*_{vol}$ defined by Eqs.~(\ref{Eq:idealVolume1})-(\ref{Eq:idealVolume2}).
$R^*_0$ values have been fitted so as to provide the best agreement 
with the curves from the MC simulations in terms of absolute Coulomb energy values. 
Note that the radii thus obtained agree very well with the ones anticipated 
from radial distribution functions of the clusters (not shown here). 
We might anticipate  some deviations in the absolute values of $U^*_{MC}$ and $U^*_{surf}$ 
(respectively  $U^*_{vol}$) 
since the clusters investigated here are not perfectly spherical, although some of them might have nearly icosahedral symmetry, but we will see below that deviations 
(when they exist) are very small in most of the cases investigated here.

For $N\le 147$, the fraction of charged particles located at the cluster surface always exceeds $\sim 0.7$ (see Fig.~\ref{fig-Fsurf}), so that a majority of the charged particles are located at the cluster surface. 
We find that these clusters behave as ideal electrical spherical conductors at solid-state temperatures where we can see that the Coulomb energy $U^*_{MC}$ closely follows the curves calculated for a sphere uniformly charged at its surface (dashed cyan curves). 
The ideal curves were obtained with $R_0^* = 1.98$, $2.50$, and $2.69$ for clusters with $N = 55$, $100$, and $147$, respectively (see Figs~\ref{fig-Ecoul}a-c). 
We observe deviations from the ideal behaviour for A$_{147}^{74+}$ ($q^*=0.5$, $Q^*=37$) when $r^*=2-3$ (close to the surface), which we assume are related to distortions from sphericity for these clusters because of the higher value of the total charge $Q^*$. 
At liquid-state temperatures (red lines in Fig.~\ref{fig-Ecoul}), 
the clusters display a behaviour between spherical conductors and spheres uniformly charged in volume. 
The Coulomb energy curve is close to the spherical conductor case for 
$r^*< 2$ where the potential energy is roughly constant (as expected for a conductor).   
Note that the constant is either equal to the one obtained at the solid-state temperature (A$_{100}^{50+}$) or slightly lower than this reference value 
(A$_{55}^{30+}$ and A$_{147}^{74+}$). 
For $r^*\ge 2$, the curves then depart from the ideal constant of the spherical 
conductor, but they nonetheless do not coincide with the ideal law of a sphere 
uniformly charged in its volume (dashed red curves which always stand below the red solid curves). 
The deviations from the ideal spherical conductor might therefore again be  
related to deviations from sphericity of the clusters at liquid-state temperatures, reflecting  more diversity of clusters conformations and shapes in order to accomodate the charges at the surface at these temperatures. 

The main effect of the addition of polarisation in the interactions between the particles in the MCPT calculations 
is to slightly decrease the number of charged particles located at the surface, as discussed in the previous section and shown in Fig.~\ref{fig-Fsurf}, or in other words  
the charged particles are slightly more prone to occupy the interior of the cluster. 
The dotted red curves in Fig.~\ref{fig-Ecoul}(a-c) of the Coulomb energy 
obtained at liquid-state temperatures of the clusters reveal that 
the smaller clusters A$_{55}^{30+}$ and A$_{100}^{50+}$ have intermediate behaviours between the two ideal cases, ie spherical conductor and sphere uniformly charged in volume, 
while A$_{147}^{74+}$ clusters depart less from the ideal case of the spherical conductor. 

Interestingly,  the electrostatic energy curves $U^*_{MC}$ for clusters simultaneously carrying 
positive and negative charged particles, as reported in Figure~\ref{fig-Ecoul}d for A$_{147}^{60+14-}$, 
now show that these clusters behave like ideal spherical conductors 
for both solid-state  and liquid-state temperatures. 
This certainly results from the fact that positively charged particles now, on average, 
solvate the negatively charged particles embedded within the volume of the clusters, 
thus reducing the apparent total charge in the interior of the cluster. 
The remaining positively charged particles located at the clusters surface then 
provide the final electrical thin layer of the conductor. 
These conclusions are not changed when including polarisation (red dotted curve). 

\begin{figure}
\epsfig{file=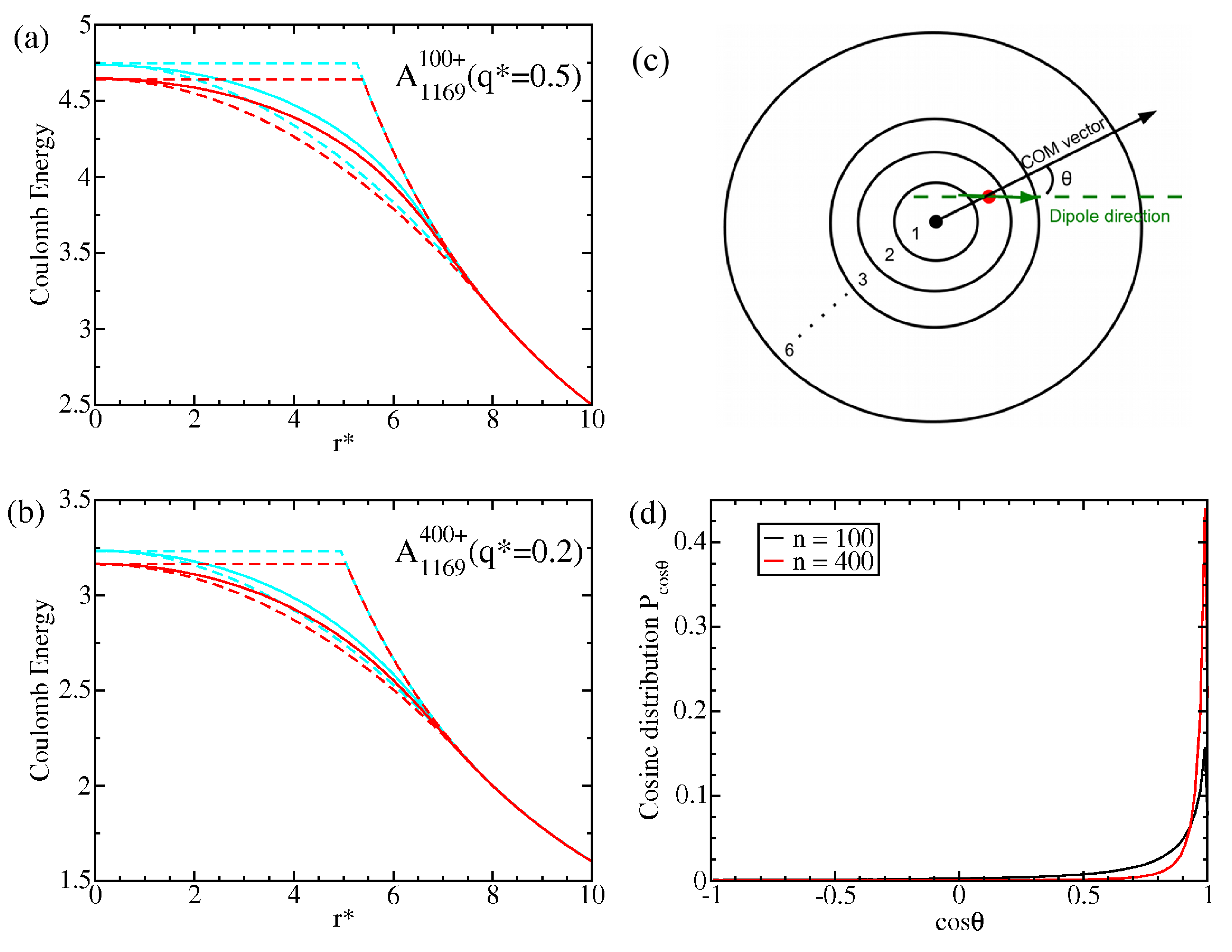,width=1.0\textwidth,angle=0}
\caption{}
\label{fig-A1169}
\end{figure}

When the cluster size is increased up to $N=1169$, the amount of charged particles located inside the cluster increases (see Fig.~\ref{fig-Fsurf}). 
For A$_{1169}^{100+}$ and A$_{1169}^{400+}$ clusters the fraction of charged particles located at the surface is in the range $0.4-0.6$, thus providing a balance between the number of charged particles located at the cluster surface and inside the cluster. 
The Coulomb energies $U^*_{MC}$ for A$_{1169}^{100+}(q^*=0.5)$ and A$_{1169}^{400+}(q^*=0.2)$ are plotted in Fig.~\ref{fig-A1169}a-b and are again compared to the ideal Coulomb energies $U^*_{surf}$ and $U^*_{vol}$. 
One can immediately see that the curves extracted from the MC calculations 
follow closely the case of spheres uniformly charged in volume over the 
whole range of distances $r^*$. 
This behaviour is observed at both solid- and liquid-state temperatures. 

We thus find three general behaviours for the charged clusters investigated here: 
\begin{enumerate}
\item The smaller A$_N^{n+}$ clusters ($N\le 147$) rather behave as ideal spherical conductors at solid-state and liquid-state temperatures. 
This is consistent with the fact that the charged particles are predominantly located at these clusters surface for this size range. 
Including polarisation does not change these conclusions. 
\item Mixed clusters A$_N^{n+m-}$ composed of $n$ positive and $m$ negative charged particles behave as ideal spherical conductors, at both solid- and liquid-state temperatures. 
In such clusters, the negative charged particles are more prone  to reside inside the clusters than the positive particles, while the positively charged particles make a compromise between solvating the negative particles within the cluster and reaching the cluster surface. 
As a result, these clusters appear as if they would carry their excess charge at the surface. 
\item The larger A$_N^{n+}$ clusters ($N = 1169$)  behave as ideal spheres uniformly charged within their volume, although the number of charged particles located at the surface roughly equals the number of charged particles embedded inside the cluster.
Also note that locating half the number of charged particles at the droplets surface is not sufficient for the droplets to experiment some conductor-like behaviour.  
This is now clearly different from Konermann's observations\cite{EAhadi2010} on a droplet composed of roughly the same number of particles (1248 in their study).
The main difference between our two studies is the absence of a permanent dipole moment attached to the particles in our droplets, as well as the absence of hydrogen bonds between the particles. 
There is thus no ``alignment'' of dipole moments, and therefore no charge-dipole induced orientation of these particles in our case. 
Note that we conceive that  A$_{1169}^{n+m-}$ would  favour the transfer of the excess charge onto the cluster surface, following the conclusions from point 2 above. 
\end{enumerate}

Charged particles in our model do not carry permanent dipoles but they can carry induced dipoles resulting from polarisation effects. 
We report in Fig.~\ref{fig-A1169}d the average distribution of $\cos\theta$ between the induced dipoles and the vector joining the cluster center of mass 
for A$_{1169}^{100+}(q^*=0.5)$ and A$_{1169}^{400+}(q^*=0.2)$ clusters at liquid-state temperatures (See pattern in Fig.~\ref{fig-A1169}c for an illustration). 
 These distributions are calculated {\it a posteriori} and are averaged over $2\times 10^3$ Monte Carlo configurations initially sampled without polarisation.
The distributions are peaked at about $\cos\theta\approx 1$ but their tail can spread up to $\cos\theta\lesssim 0$ (ie, $\theta \gtrsim 90^\circ$) for A$_{1169}^{100+}(q^*=0.5)$ clusters. 
The larger the number $n$ of charged particles, the sharper the distribution peaks and the shorter the extent of these distributions.
This reflects the differences in the induced electric fields created by the charged particles (due to different charge distributions within the cluster) which result into more directionality of the induced dipoles as $n$ increases. 
A finer analysis of the structure in layers of these clusters (not shown here)  also reveals that the induced dipoles are stronger and more aligned along the center of mass vector for particles farther from the cluster center of mass.  

\section{Conclusion}\label{sec-conclusion}

Monte Carlo simulations have been performed on A$_N^{n+}$ clusters composed of $N$ particles ($55\le N\le 1169$) among which $n$ are positively charged with charge $q^*$, the cluster having a total net charge $Q^*=nq^*$. 
We have also investigated A$_N^{n+m-}$ clusters carrying $n$ positively and $m$ 
negatively charged particles for a total net charge $Q^*=(n-m)q^*$. 
These clusters have been modelled with a simple and generic representation of the 
interactions between individual particles (Lennard-Jones and Electrostatic 
interactions, including polarisation), in dimensionless units. 
The temperature range of the simulations goes from solid-like to liquid-like states, encompassing the melting transition. 
We have investigated the localisation of the charged particles within the clusters, 
i.e. at the surface or embedded within the volume of the clusters, 
as well as the behaviour of these charged clusters in terms of ideal conductors 
or ideal spheres uniformly charged in volume. 
Both questionings stem from Electrosprayed droplets produced in Mass Spectrometry 
experiments (ESI-MS) for which these properties might be pivotal for the final 
production of the ions in the gas phase. 

For the charged clusters  investigated here, with a total net positive charge $Q^*$, the charged particles tend to be predominantly located at the clusters surface at finite temperature. 
We found that an increase of $Q^*$ at fixed $n$ or a decrease of $n$ at fixed $Q^*$  favoured the migration of the positive particles to the cluster surface although complete occupancy of the surface sites was found out of reach for the larger  $A_{1169}^{n+}$ clusters, even when the number of charged particles was larger than the number of surface sites.
For the smaller clusters $N \leq 147$, low temperatures were found to favour the exploration of regions around the global minimum attraction basins.
These regions are entropically unfavoured as temperature increases, especially in the melting- and liquid-state regimes of interest for experiments such as ESI.
The localization of charged particles on the cluster surface still remains favoured in these regimes, with more than 70-80\% occupancy for small clusters. 
Inclusion of polarisation slightly reduces the filling up of the cluster surface with charged particles and, interestingly, the study of charged clusters composed of both positive and negative particles revealed that positive particles were subject to two competing localization mechanisms: filling up of surface sites  and solvation of negative particles.

Considering the electrical behaviour of these charged clusters, we have found two main behaviours. 
The smaller A$_N^{n+}$ clusters ($N\le 147$)  behave as ideal spherical conductors at solid-state and liquid-state temperatures, which is consistent with the charged particles being predominantly located at the cluster surface. 
Interestingly, for the same range of clusters sizes ($N = 147$ here), 
mixed A$_N^{n+m-}$ clusters composed of $n$ positively and $m$ negatively 
charged particles display a reduced number of charges located at the surface:  
negative charged particles are more prone  to reside inside the clusters than the positive particles, while the positively charged particles make a compromise 
between solvating the negative particles within the cluster and reaching the cluster 
surface. 
Despite the reduced number of charges located at the surface of A$_N^{n+m-}$, 
these clusters behave as ideal spherical conductors, where the excess charge 
is located within a thin layer at the clusters surface. 
On the contrary, the larger A$_N^{n+}$ clusters ($N = 1169$ here) 
behave as ideal spheres uniformly charged within their volume, although the number 
of charged particles located at the surface roughly equals the number of charged 
particles embedded inside the cluster.

This last result is clearly different from Konermann's results\cite{EAhadi2010} on a droplet composed of roughly the same number of particles, i.e. 
1248 microscopic water molecules (SPC model) and 10 Na$^+$ ions. 
The main difference between our two studies is the level of representation, 
i.e. microscopic representation of the molecules and atomic ions in Ref\cite{EAhadi2010} and a coarse-grained representation in our present investigation. 
In our representation, there is no permanent dipole moment attached to the particles  as well as no hydrogen bonds between the particles, contrary to the microscopic representation of Konnerman and coworker. 
There is thus no possibility for ``alignment'' of dipole moments, 
and therefore no possibility for charge-dipole induced orientation of these particles in our droplets, 
which is the main argument put forward by Konermann and coworker in Ref.\cite{EAhadi2010} for explaining the excess charge at the surface of their droplets. 
Note that the induced dipole moments, taken into account in our calculations on 
the A$_{1169}^{n+}$ clusters (and not considered in Ref.\cite{EAhadi2010}), are not large enough to induce similar charge-dipole alignments. 
As a result, the only A$_N^{n+}$ droplets that behave as ideal conductors are the ones composed of at most about 100 particles, because the charged particles are 
predominantly located at the clusters surface. 
The larger droplets of more than 1000 particles behave as spheres uniformly 
charged in volume, although 
roughly half of the charged particles are located at the clusters surface. 
Note that our current coarse-grained representation of the solvent in the droplets 
is very similar to the coarse-grained models of solvent developed and recently 
reviewed by Klein {\it et al.}~\cite{XHe2010}, 
i.e. chargeless and dipoless coarse grains which can be mapped to up to  four individual microscopic water molecules. 
This is thus a reasonable model for treating coarse-grained droplets, and the results obtained here are therefore deemed representative of charged droplets containing coarse-grained water. 

Our present findings on charged droplets should be taken into consideration when investigating charged droplets produced by ESI. 
In particular, the common assumption according which positive charges lie on the droplet surface should be taken with caution since the thermodynamic and electrical properties of the droplet may strongly affect the propensity of positive charges for surface localisation, as shown here. 
It is not clear how the electrical properties found here on the smaller charged 
droplets being ideal conductors while larger charged droplets are spheres uniformly charged in volume translate for the fragmentation mechanisms of these droplets in ESI experiments. 
We are currently investigating the consequences of these findings on the fragmentation mechanisms and possible correspondence with the Charge Residue Model (CRM) or the Ion Evaporation Model (IEM).
We are also investigating the effects of applying an external electric field.

More realistic interaction potential models should also be developped, 
typically inhomogeneous spherical particles (with different $u$ and $\sigma$ parameters), non-spherical particles, including patchy particles and in particular patchy proteins, and possibly going towards coarse-grained proteins embedded in droplets using the Martini or other modern coarse-grained models. 
This is another direction we are exploring.

\section{Acknowledgments}
Dr Mark A. Miller is gratefully acknowledged for helpful assistance and strong interest in this work.
We are grateful to Prof. Daan Frenkel and Dr. Florent Calvo for fruitful discussions.
We also acknowledge reviewers for suggesting the discussion on electrical properties of charged clusters.
The IDRIS national computer center and the ROMEO computer center of Champagne-Ardenne are acknowledged for computer time.
This work was supported by G\'enop\^ole-Evry through a Post-Doctoral fellowship (DAB), and by a Partenariat Hubert Curien Alliance Program. 

\section{Appendix: Coulomb energies in reduced units}

The Coulomb energy $U_{vol}$ between a probe particle with charge $q_{probe}$ and the electrostatic potential $V_{vol}$ created by a sphere of radius $R_0$ uniformly charged in volume with density of charge $\rho_Q=Q/V$ ($Q$ being the total charge of the sphere and $V$ its volume) is $U_{vol}(r) = q_{probe}V_{vol}(r)$ where
\begin{eqnarray*}
r < R_0 & : & V_{vol}(r) = \frac{\rho_Q}{2\epsilon_0}\left(R_0^2-\frac{r^2}{3}\right) \\
r > R_0 & : & V_{vol}(r) = \frac{\rho_Q R_0^3}{3\epsilon_0 r}
\end{eqnarray*}
In reduced units, $r^* = r/\sigma$ for distances, $U^*=U/u$ for energies, and $q^*=q/\sqrt{4\pi\epsilon_0\sigma u}$ for charges, this leads to
\begin{eqnarray*}
r < R_0 & : & V_{vol} =  \frac{\rho_Q^*}{2}\left(R_0^{*2}-\frac{r^{*2}}{3}\right)\sqrt{\frac{4\pi u}{\sigma\epsilon_0}}\\
r > R_0 & : & V_{vol} = \frac{\rho_Q^* R_0^{*3}}{3r^*} \sqrt{\frac{4\pi u}{\sigma\epsilon_0}}
\end{eqnarray*}
To properly define $V_{vol}^*$  in reduced units, we use the simple and general expression of the electrostatic potential created by a charge $q$ 
\begin{eqnarray*}
V_{vol} = \frac{q}{4\pi\epsilon_0\,r} = \frac{q^*}{r^*}\sqrt{\frac{u}{4\pi\epsilon_0\sigma}} = V_{vol}^*\sqrt{\frac{u}{4\pi\epsilon_0\sigma}}
\end{eqnarray*}
when posing $V^*_{vol} = \frac{q^*}{r^*} = V_{vol}\sqrt{\frac{4\pi\epsilon_0\sigma}{u}}$. 
We then obtain
\begin{eqnarray*}
r^* < R_0^* & : & V^*_{vol} = \frac{4\pi\rho_Q^*}{2}\left(R_0^{*2}-\frac{r^{*2}}{3}\right)\\
r^* > R_0^* & : & V^*_{vol} = \frac{4\pi\rho_Q^* R_0^{*3}}{3r^*}.
\end{eqnarray*}
In reduced units the Coulomb energy writes $U_{vol} = q_{probe}V_{vol} = u\,q_{probe}^*V^*_{vol} = u\,U^*_{vol}$ when naturally posing $U^*_{vol} = q_{probe}^*V^*_{vol} = U_{vol}/u$ since $U_{vol}$ is an energy.
We eventually find 
\begin{eqnarray}
r^* < R_0^* & : & U^*_{vol} = \frac{4\pi q_{probe}^*\rho_Q^*}{2}\left(R_0^{*2}-\frac{r^{*2}}{3}\right)\\
r^* > R_0^* & : & U^*_{vol} = \frac{4\pi q_{probe}^*\rho_Q^* R_0^{*3}}{3r^*}.
\end{eqnarray}

For a spherical conductor with density of charge $\sigma_Q=Q/S$ ($S$ being the surface of the sphere) the derivation of the Coulomb energy $U^*_{surf}$ is equivalent. 
The expression of the electrostatic potential $V_{surf}$ is 
\begin{eqnarray*}
r < R_0 & : & V_{surf} = \frac{\sigma_Q R_0}{\epsilon_0}\\
r > R_0 & : & V_{surf} = \frac{\sigma_Q R_0^2}{\epsilon_0 r}
\end{eqnarray*}
which leads to 
\begin{eqnarray*}
r^* < R_0^* & : & U^*_{surf} = 4\pi q_{probe}^*\sigma_Q^*R_0^*\\
r^* > R_0^* & : & U^*_{surf} = \frac{4\pi q_{probe}^*\sigma_Q^* R_0^{*2}}{r^*}.
\end{eqnarray*}



\begin{thebibliography}{39}
\bibitem[{Not()}]{NoteMCPT}
\bibinfo{note}{The optimum overlap between adjacent distributions is often
  considered to be comprised between 30\%-40\% and 60\%-70\%. However, for
  large clusters, and especially for $A_{1169}^{n+}$ clusters, the melting peak
  is so narrow that higher overlaps are needed to resolve the melting peak.}
\end{thebibliography}

\clearpage
\Kaption{1}{Distributions of $\Delta V^*$, the energy difference between the local minima explored along the MCPT simulations and the corresponding global minimum for selected A$_{55}^{30+}$ and A$_{147}^{n+}$ clusters at solid-state (cyan curves), melting (black curves), and liquid-state (red curves) temperatures. 
For clarity, the breadths of A$_{55}^{30+}$ distributions have been multiplied by 2.6.}

\Kaption{2}{Average fraction $F_{surf}$ of charged particles located at the cluster surface as a function of the total cluster charge $Q^*$ for A$_N^{n+}$ ($Q^*=nq^*$) clusters (open symbols) and A$_N^{n+m-}$ ($Q^*=(n-m)q^*$) clusters (closed symbols) of size (a) N=55, (b) N=100, (c) N=147, and (d) N=1169. 
The symbols correspond to fractions obtained at the melting temperature. 
The error bars indicate the range of values spanned by $F_{surf}$ for the temperatures of the Parallel Tempering simulations. 
Simulations without  polarisation (black symbols), with polarisation (red symbols). 
Note that only charged particles were chosen polarisable for A$_{147}^{n+}$ and A$_{147}^{n+m-}$ clusters.}

\Kaption{3}{Average Coulomb energy between a probe particle with charge $|q^*|$ and the charged particles composing selected (a) A$_{55}^{30+}$, (b) A$_{100}^{50+}$, (c) A$_{147}^{74+}$, and (d) A$_{147}^{60+14-}$ clusters.
These energies are determined at solid-state temperature without polarisation (cyan solid curves) and liquid-state temperature without polarisation (red solid curves) and with polarisation (red dotted curves).
Comparison with the ideal Coulomb energies $U^*_{surf}$ and $U^*_{vol}$ are also reported (dashed curves).
See text for the formula of the ideal Coulomb energies $U^*_{surf}$ and $U^*_{vol}$.}

\Kaption{4}{Average Coulomb energy between a probe particle with charge $|q^*|$ and the charged particles composing selected (a) A$_{1169}^{100+}$ and (b) A$_{1169}^{400+}$ clusters.
Comparison with the ideal Coulomb energies $U^*_{surf}$ and $U^*_{vol}$ are also reported (dashed curves).
(c) Pattern that defines the angle $\theta$ used for plotting cosine distributions $P_{\cos\theta}$ of large clusters. 
(d) Average $P_{\cos\theta}$ distributions for the estimated induced dipoles on A$_{1169}^{100+}(q^*=0.5)$ (black curves) and A$_{1169}^{400+}(q^*=0.2)$ (red curves).}

\clearpage \newpage 

\clearpage

\clearpage

\end{document}